1# Observation of SOMO-HOMO inversion in a polycyclic conjugated hydrocarbon

Shantanu Mishra[1], Manuel Vilas-Varela[2], Shadi Fatayer[3], Florian Albrecht[1], Diego Peña[2,4], and Leo Gross[1]

[1]IBM Research Europe – Zurich, 8803 Rüschlikon, Switzerland.

[2]Center for Research in Biological Chemistry and Molecular Materials (CiQUS) and Department of Organic Chemistry, University of Santiago de Compostela, 15782 Santiago de Compostela, Spain.

[3]Applied Physics Program, Physical Science and Engineering Division, King Abdullah University of Science and Technology (KAUST), 23955-6900 Thuwal, Kingdom of Saudi Arabia.

[4]Oportunius, Galician Innovation Agency (GAIN), 15702 Santiago de Compostela, Spain.

Corresponding authors: Shantanu Mishra (SHM@zurich.ibm.com), Diego Peña (diego.pena@usc.es) and Leo Gross (LGR@zurich.ibm.com).**Abstract:** We report the generation of a non-benzenoid polycyclic conjugated hydrocarbon, which consists of a biphenyl moiety substituted by indenyl units at the 4,4′ positions, on ultra-thin sodium chloride films by tip-induced chemistry. Single-molecule characterization by scanning tunneling and atomic force microscopy reveals an open-shell biradical ground state, with a peculiar electronic configuration wherein the singly-occupied molecular orbitals (SOMOs) are lower in energy than the highest occupied molecular orbital (HOMO).There is currently an immense interest and progress in organic radicals, and in particular in open-shell polycyclic conjugated hydrocarbons. This is driven by both advancements in on-surface chemistry for synthesis and stabilization of reactive species,[1] and the fundamental insights that organic radicals provide into chemical reactions, many-body nature of electronic wavefunctions and theoretical models in quantum magnetism, with implications for spintronic and optoelectronic technologies.[2–4] In a molecular orbital picture, a closed-shell molecule (Figure 1a) is represented by a series of doubly-occupied molecular orbitals leading up to the HOMO, and a series of empty molecular orbitals beginning with the lowest unoccupied molecular orbital (LUMO). HOMO and LUMO are the frontier molecular orbitals, which to a great extent govern the chemical reactivity, electronic transport and low-energy optical transitions of a molecule. Conversely, for the majority of open-shell molecules (for simplicity, we first consider a monoradical; Figure 1b), the frontier molecular orbital is a singly-occupied molecular orbital (SOMO),[5,6] whose one spin level (spin up or spin down) is occupied, and the corresponding opposite-spin level, referred to as SUMO, is empty. Interestingly, there have been reports on radicals with an unusual electronic configuration, where the energy of the SOMO lies below one or more doubly-occupied molecular orbitals (Figure 1c,d).[7] This phenomenon is referred to as SOMO-HOMO inversion (SHI). SHI has been associated with enhanced radical stability, for example, improved photostability and enhanced quantum yield of luminescent radicals,[8] and generation of high-spin states upon radical oxidation that could be exploited for the development of organic ferromagnetic materials.[9] Furthermore, chiral conjugated SHI molecules are attracting interest given the potential of designing functionalities such as circularly-polarized luminescence and chirality-induced spin selectivity.[10] SHI radicals have also been utilized as key intermediates in chemical reactions, such as in amination of imidates and amidines,[11] and in photocatalytic allylation.[12]

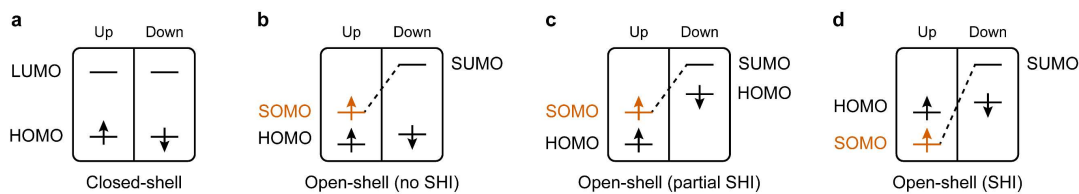

**Figure 1.** Schematic representation of the electronic configuration of (a) closed-shell molecule, (b) open-shell molecule without SHI, (c) open-shell molecule with partial SHI, wherein the energy of the SOMO lies in between the spin-up and spin-down HOMO levels, and (d) open-shell molecule with SHI, wherein the energy of the SOMO is lower than both the spin levels of HOMO.

The topic of SHI in organic radicals dates back to the 1987 study of Awaga et al., who performed spin-unrestricted calculations on galvinoxyl radical and demonstrated the existence of partial SHI in the molecule (Figure 1c).[13] Experimentally, the first organic radical exhibiting SHI was a tetrathiafulvalene derivative substituted with imino pyrolidine and piperidine-1-oxyl, synthesized by Sugimoto et al. in 1993.[14] Since then, SHI has been experimentally found in a number of molecular systems, such as aromatic heterocycles substituted with aminoxyl radicals,[15–19] nitroxyl metalladithiolates,[20,21] carboxy-tetrathiafulvalene crystals,[22] and metalloporphyrins and metallophthalocyanines.[23] More recently, SHI has also been experimentally shown to occur in distonic radical ions,[24,25] sulfur- and nitrogen-containing helicene-based neutral radicals and radical ions,[26,27] carbazole-based radical ions,[10,28] triarylmethyl-radical-based donor-acceptor frameworks,[8,29] spiro-conjugated donor-acceptor charged frameworks[30] and spiro-fused diarylaminyl radicals.[31] Theoretically, SHI has been shown to occur in biologically important radicals such as singly-oxidized nucleobases and DNA base pairs,[32,33] singly-reduced radicals of diatomic molecules such as $O_2$ and BN,[33] triplet carbene-based frameworks,[34] triplet cyclopentane-1,3-diyl-based diradicals,[35] oxidized azulene derivatives,[36] donor-acceptor conjugated polymers,[37] napthalene diimides substituted with pyridines,[38] helical π-systems,[39] triplet polycyclic hydrocarbons[40] and porphyrin oligomers.[41] Abella et al. conducted a detailed theoretical study of the mechanism of SHI in organic radicals, which were conceptually generated by one-electron oxidation of the corresponding closed-shell parent compounds that contained an additional electron.[38] Their analysis revealed three key criteria for SHI to occur. First, a strongly-localized HOMO of the closed-shell parent results in a strong self-Coulomb repulsion, that is, repulsion between electrons in the HOMO. Upon ionization, relieving of the self-Coulomb repulsion leads to a notable drop in the energy of this orbital. Second, spatially-disjoint HOMO and HOMO–1 of the closed-shell parent results in a weak Coulomb repulsion between the two orbitals, and consequently, the energy of the latter orbital drops far less upon ionization. Third, a small gap between HOMO–1 and HOMO of the closed-shell parent promotes this energetic crossover upon ionization.

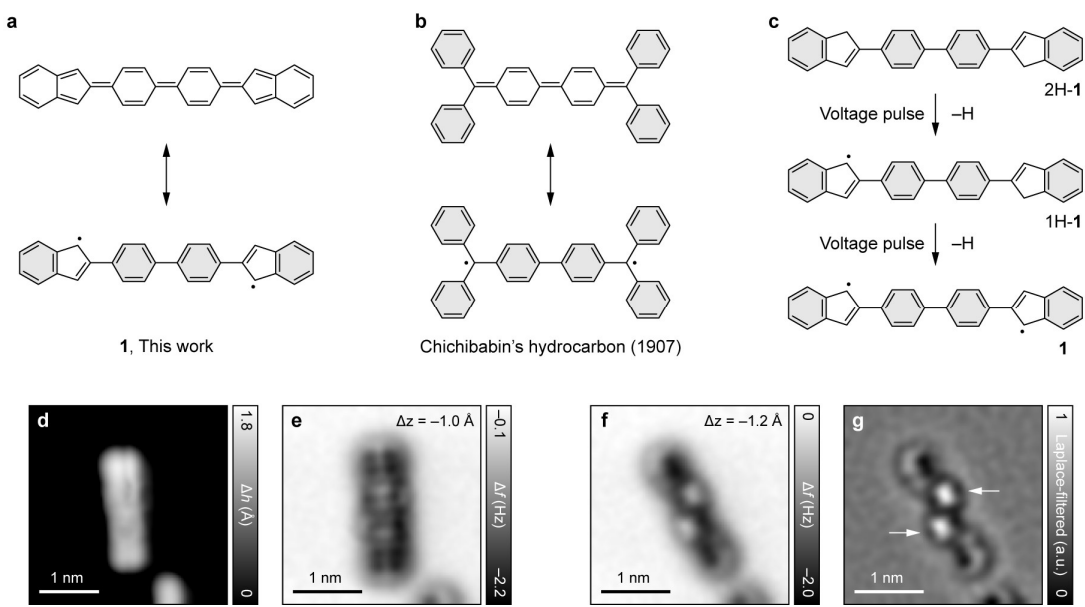

**Figure 2.** Structural characterization of **1**. (a, b) Closed-shell Kekulé (top) and open-shell non-Kekulé (bottom) resonance structures of **1** (a) and Chichibabin's hydrocarbon (b). (c) Scheme of on-surface generation of **1** through voltage-pulse-induced dehydrogenation of 2H-**1**. Compound 1H-**1** denotes the intermediate monoradical species after removal of one hydrogen atom from 2H-**1**. Gray filled rings denote aromatic π-sextets. (d) In-gap STM image of **1** ($V$ = 0.2 V and $I$ = 0.5 pA; $V$ and $I$ denote the bias voltage and tunneling current, respectively). $\Delta h$ denotes the tip height. The species at the bottom-right of the scan frame is a 2H-**1** molecule. (e) Corresponding AFM image of **1**. $\Delta f$ and $\Delta z$ denote the frequency shift and the tip-height offset with respect to the STM setpoint, respectively. Positive and negative values of $\Delta z$ denote tip approach and retraction from the STM setpoint, respectively. Here, **1** is changing its adsorption site under the influence of the tip during scanning, which causes the apparent bisection of the molecule along its long axis. (f, g) AFM image (f) and the corresponding Laplace-filtered version (g) of the same molecule as in (e), with its adsorption orientation changed by ~20°. a.u. denotes arbitrary units. The movement of **1** is now hindered by the 2H-**1** species at the bottom. The two bright features at the center of **1** that are indicated with arrows in (g) correspond to the strongly-tilted hexagonal rings of the biphenyl moiety. The indenyl units also tilt, but to a lesser degree. STM setpoint for AFM images: $V$ = 0.2 V and $I$ = 0.5 pA on bilayer NaCl.

To date, the experimental existence of SHI has been established through bulk spectroscopic measurements, such as absorption and photoelectron spectroscopies, cyclic voltammetry and electron spin resonance. However, SHI has not been detected at the single-molecule level, and SHI has never been observed in a neutral polycyclic conjugated hydrocarbon. Here, we report the generation of a non-benzenoid open-shell polycyclic conjugated hydrocarbon **1** ($C_{30}H_{20}$; Figure 2a), which exhibits SHI. Compound **1** is structurally related to Chichibabin's hydrocarbon ($C_{38}H_{28}$; Figure 2b),[42,43] where the terminal diphenylmethylene substituents in Chichibabin's hydrocarbon are replaced by indenyl substituents in **1**. Compared to their respective closed-shell structures, **1** and Chichibabin's hydrocarbon gain four and two aromatic sextets in their open-shell structures, respectively. Therefore, a larger biradical open-shell character is expected for **1** compared to Chichibabin's hydrocarbon. Compound **1** was generated from the stable dihydro precursor 2H-**1** ($C_{30}H_{22}$; Figure 2c), which was synthesized in solution by double Suzuki coupling (see Supporting Information). Single molecules of 2H-**1** were deposited on a single-crystal Au(111) surface that was partially covered by two-monolayers-thick (denoted bilayer) NaCl films (Figure S1–S3), and housed inside a combined scanning tunneling microscopy (STM) and atomic force microscopy (AFM) apparatus operating under ultra-high vacuum and at a temperature of 5 K. Voltage pulses ranging between 5.8 and 6.4 V were applied to individual 2H-**1** molecules by the tip of the STM/AFM system, resulting in the homolytic cleavage of the C($sp^3$)-H bond at each of the pentagonal rings, thereby leading to the generation of **1** (Figure 2c). In the main text, we focus on the characterization of **1** on bilayer NaCl/Au(111). For generation and characterization of **1** on Au(111), see Figure S4. Figures 2d,e present STM and AFM images of **1** generated after a voltage-pulse sequence. The images reveal an apparent bisection of **1** along the long molecular axis, which results from the movement of **1** between different adsorption sites under the influence of the microscope tip.[5] The adsorption of **1** can be stabilized in the vicinity of a defect, adsorbate or step edge. Figure 2f shows the AFM image of the same molecule after its lower part moved adjacent to a 2H-**1** species, thereby leading to stable adsorption. The corresponding Laplace-filtered AFM image (Figure 2g) resolves the terminal indenyl units, along with two bright features (that is, with higher frequency shift due to stronger repulsive forces) in the central part that correspond to the strongly-tilted hexagonal rings of the biphenyl moiety.[44]

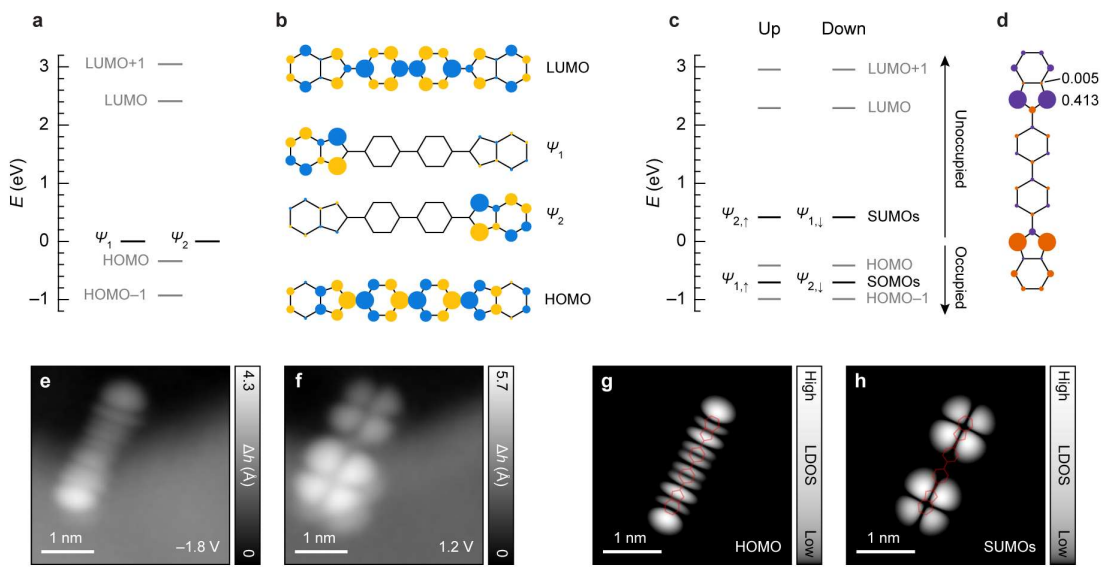

**Figure 3.** Electronic characterization of **1**. (a) Tight-binding energy spectrum of **1**. Zero of the energy axis is set to match the energy of the degenerate states $\Psi_1$ and $\Psi_2$. The nearest-neighbor hopping parameter is 2.7 eV. (b) Tight-binding wavefunctions of HOMO, $\Psi_1$, $\Psi_2$ and LUMO. Size and color of the circles denote the amplitude and phase of the wavefunctions. (c) Mean-field Hubbard energy spectrum of **1** in the open-shell singlet state. Zero of the energy axis is set to lie in between the HOMO and SUMOs. States above and below zero energy are unoccupied and occupied, respectively. The on-site Coulomb repulsion is ~4 eV. (d) Spin-polarization plot of **1** in the open-shell singlet state, expressed as the difference in mean populations of spin-up and spin-down electrons at each carbon atom. Size and color of the circles denote the value and sign of spin polarization, respectively. The numbers denote absolute values of spin polarization. (e, f) STM images of **1**, immobilized at a gold step edge overgrown by NaCl, acquired at the positive (e) and negative (f) ion resonances ($I$ = 0.1 pA). (g, h) Constant-height mean-field Hubbard local density of states (LDOS) maps of HOMO (g) and SUMOs ($|\Psi_{1,\downarrow}|^2 + |\Psi_{2,\uparrow}|^2$; h) of **1**, calculated at a height of 5.6 Å above the molecular plane and shown in a logarithmic color scale. AFM image of the molecule in (e) and (f) is shown in Figure S9.

To describe the electronic structure of **1**, we start by performing nearest-neighbor tight-binding calculations. Our tight-binding results are supported by density functional theory calculations shown in Figure S5. Figure 3a shows the tight-binding energy spectrum of **1**, where the salient features correspond to a pair of degenerate states labeled $\Psi_1$ and $\Psi_2$, whose wavefunctions are strongly localized on either of the indenyl units, and the HOMO and LUMO states that are delocalized over the molecular framework (Figure 3b). Of note is the close energetic proximity of HOMO to $\Psi_1/\Psi_2$, and a much larger separation between LUMO and $\Psi_1/\Psi_2$. To describe the formation of local magnetic moments and magnetic exchange in **1**, we include electronic correlations via the Hubbard approximation, where an on-site Coulomb repulsion term is included in the tight-binding Hamiltonian – the mean-field Hubbard model. Previous works have shown the Hubbard approximation to be consistent with spin-polarized density functional theory calculations.[45,46] The mean-field Hubbard solution leads to spin polarization of **1**, with nearly degenerate open-shell singlet and triplet states (Figure S6), as also predicted by our density functional theory calculations (Figure S5). Figure 3c shows the mean-field Hubbard energy spectrum of **1** in the open-shell singlet state (our experimental results are also consistent with the triplet state of **1**, see Figure S6), with the corresponding spin polarization plot shown in Figure 3d. While there are negligible changes in the energies of the HOMO and LUMO from the tight-binding case, the degeneracy of $\Psi_1$ and $\Psi_2$ is lifted by spin polarization, along with the opening of a Hubbard gap in the energy spectrum. At this stage, SHI is already established in **1**. The energies of the SOMOs, that is, the occupied spin levels of $\Psi_1$ and $\Psi_2$, are lowered to the extent that these states shift below the HOMO. This is likely aided by the strongly-localized character of $\Psi_1$ and $\Psi_2$ (Figure 3b) that promotes substantial lowering of the energy of these states upon spin polarization, along with the previously noted small tight-binding gap between HOMO and $\Psi_1/\Psi_2$ (Figure 3a) that promotes such a crossover. Concurrently, the tight-binding gap between LUMO and $\Psi_1/\Psi_2$ is large enough such that the SUMOs, that is, the unoccupied spin levels of $\Psi_1$ and $\Psi_2$, remain as the frontier unoccupied states upon spin polarization. The existence of SHI in **1** is experimentally confirmed by STM imaging at the ionic resonances. The positive (negative) ion resonance denotes transition between the neutral and cationic (anionic) states of a molecule, which corresponds to electron detachment from (attachment to) the occupied (unoccupied) frontier molecular orbitals. STM images of **1** acquired at the positive (Figure 3e) and negative (Figure 3f) ion resonances show orbital densities that concur with the calculated local density of states maps of the HOMO (Figure 3g) and SUMOs (Figure 3h), respectively. The observation of HOMO and SUMOs as the frontier orbitals thus proves SHI in **1**. SHI is also observed for the monoradical species 1H-**1** (Figure S7 and S8). Whereas SHI manifests in **1** in the form of two energetically degenerate SOMOs that lie below the HOMO, in 1H-**1** a single SOMO lies below the HOMO. The latter scenario, schematically depicted in Figure 1, has been previously observed for several molecular systems.

In conclusion, we generated a non-benzenoid polycyclic conjugated hydrocarbon **1** on bilayer NaCl/Au(111) by scanning-probe-based atom manipulation. Compound **1** consists of a biphenyl moiety substituted by indenyl units at the 4,4′ positions. Structural characterization of **1** by AFM reveals a non-planar adsorption conformation with strongly-tilted hexagonal rings of the biphenyl moiety. Electronic characterization of **1** by STM-based orbital-density imaging reveals an open-shell biradical ground state, with manifestation of the unusual phenomenon of SHI – wherein the SOMOs are lower in energy than the HOMO, which is in contrast to the majority of reported open-shell molecules. The demonstrated ability to detect SHI at the single-molecule level facilitates future experiments and applications of inducing or switching (on and off) SHI by changes in the adsorption site[47] or molecular conformation,[48] through proximity to adsorbates,[49] or by means of external fields.[50]

**Associated Content**

**Supporting Information.** Scanning probe microscopy measurements and calculations (Figures S1–S9), solution synthesis of 2H-**1** (Scheme S1), mass spectrometry of 2H-**1** (Figures S10 and S11), and experimental and theoretical methods.

**Author Information**

**Notes.** The authors declare no competing financial interest.

**Acknowledgements**

This study has received funding from the European Union project SPRING (grant number 863098), the European Research Council Synergy grant MolDAM (grant number 951519), the Spanish Agencia Estatal de Investigación (grant number PID2022-140845OB-C62), Xunta de Galicia (Centro de Investigación de Galicia accreditation 2019–2022, grant number ED431G 2019/03), the European Regional Development Fund, and the KAUST Office of Sponsored Research (grant number OSRCRG2022-503). Computational support from the Supercomputing Laboratory at KAUST is acknowledged.

**References**

(1) Clair, S.; de Oteyza, D. G. Controlling a Chemical Coupling Reaction on a Surface: Tools and Strategies for On-Surface Synthesis. *Chem. Rev.* **2019**, *119* (7), 4717–4776. https://doi.org/10.1021/acs.chemrev.8b00601.

(2) Stuyver, T.; Chen, B.; Zeng, T.; Geerlings, P.; De Proft, F.; Hoffmann, R. Do Diradicals Behave Like Radicals? *Chem. Rev.* **2019**, *119* (21), 11291–11351. https://doi.org/10.1021/acs.chemrev.9b00260.

(3) Mishra, S.; Catarina, G.; Wu, F.; Ortiz, R.; Jacob, D.; Eimre, K.; Ma, J.; Pignedoli, C. A.; Feng, X.; Ruffieux, P.; Fernández-Rossier, J.; Fasel, R. Observation of Fractional Edge Excitations in Nanographene Spin Chains. *Nature* **2021**, *598* (7880), 287–292. https://doi.org/10.1038/s41586-021-03842-3.

(4) Oteyza, D. G. de; Frederiksen, T. Carbon-Based Nanostructures as a Versatile Platform for Tunable π-Magnetism. *J. Phys.: Condens. Matter* **2022**, *34* (44), 443001. https://doi.org/10.1088/1361-648X/ac8a7f.

(5) Pavliček, N.; Mistry, A.; Majzik, Z.; Moll, N.; Meyer, G.; Fox, D. J.; Gross, L. Synthesis and Characterization of Triangulene. *Nat. Nanotechnol.* **2017**, *12* (4), 308–311. https://doi.org/10.1038/nnano.2016.305.

(6) Mishra, S.; Beyer, D.; Eimre, K.; Kezilebieke, S.; Berger, R.; Gröning, O.; Pignedoli, C. A.; Müllen, K.; Liljeroth, P.; Ruffieux, P.; Feng, X.; Fasel, R. Topological Frustration Induces Unconventional Magnetism in a Nanographene. *Nat. Nanotechnol.* **2020**, *15* (1), 22–28. https://doi.org/10.1038/s41565-019-0577-9.

(7) Gryn'ova, G.; Coote, M. L.; Corminboeuf, C. Theory and Practice of Uncommon Molecular Electronic Configurations. *Wiley Interdiscip. Rev. Comput. Mol. Sci.* **2015**, *5* (6), 440–459. https://doi.org/10.1002/wcms.1233.

(8) Guo, H.; Peng, Q.; Chen, X.-K.; Gu, Q.; Dong, S.; Evans, E. W.; Gillett, A. J.; Ai, X.; Zhang, M.; Credgington, D.; Coropceanu, V.; Friend, R. H.; Brédas, J.-L.; Li, F. High Stability and Luminescence Efficiency in Donor–Acceptor Neutral Radicals Not Following the Aufbau Principle. *Nat. Mater.* **2019**, *18* (9), 977–984. https://doi.org/10.1038/s41563-019-0433-1.

(9) Sugawara, T.; Matsushita, M. M. Spintronics in Organic π-Electronic Systems. *J. Mater. Chem.* **2009**, *19* (12), 1738–1753. https://doi.org/10.1039/B818851N.

(10) Kasemthaveechok, S.; Abella, L.; Jean, M.; Cordier, M.; Vanthuyne, N.; Guizouarn, T.; Cador, O.; Autschbach, J.; Crassous, J.; Favereau, L. Carbazole Isomerism in Helical Radical Cations: Spin Delocalization and SOMO–HOMO Level Inversion in the Diradical State. *J. Am. Chem. Soc.* **2022**, *144* (16), 7253–7263. https://doi.org/10.1021/jacs.2c00331.

(11) Zhao, R.; Fu, K.; Fang, Y.; Zhou, J.; Shi, L. Site-Specific C(sp$^3$)–H Aminations of Imidates and Amidines Enabled by Covalently Tethered Distonic Radical Anions. *Angew. Chem. Int. Ed.* **2020**, *59* (46), 20682–20690. https://doi.org/10.1002/anie.202008806.

(12) Levernier, E.; Jaouadi, K.; Zhang, H.-R.; Corcé, V.; Bernard, A.; Gontard, G.; Troufflard, C.; Grimaud, L.; Derat, E.; Ollivier, C.; Fensterbank, L. Phenyl Silicates with Substituted Catecholate Ligands: Synthesis, Structural Studies and Reactivity. *Chem. Eur. J.* **2021**, *27* (34), 8782–8790. https://doi.org/10.1002/chem.202100453.

(13) Awaga, K.; Sugano, T.; Kinoshita, M. Ferromagnetic Intermolecular Interaction in the Galvinoxyl Radical: Cooperation of Spin Polarization and Charge-Transfer Interaction. *Chem. Phys. Lett.* **1987**, *141* (6), 540–544. https://doi.org/10.1016/0009-2614(87)85077-7.

(14) Sugimoto, T.; Yamaga, S.; Nakai, M.; Ohmori, K.; Tsujii, M.; Nakatsuji, H.; Fujita, H.; Yamauchi, J. Intramolecular Spin-Spin Exchange in Cation Radicals of Tetrathiafulvalene Derivatives Substituted with Imino Pyrolidine- and Piperidine-1-Oxyls. *Chem. Lett.* **1993**, *22* (8), 1361–1364. https://doi.org/10.1246/cl.1993.1361.

(15) Nakazaki, J.; Matsushita, M. M.; Izuoka, A.; Sugawara, T. Novel Spin-Polarized TTF Donors Affording Ground State Triplet Cation Diradicals. *Tetrahedron Lett.* **1999**, *40* (27), 5027–5030. https://doi.org/10.1016/S0040-4039(99)00925-9.

(16) Izuoka, A.; Hiraishi, M.; Abe, T.; Sugawara, T.; Sato, K.; Takui, T. Spin Alignment in Singly Oxidized Spin-Polarized Diradical Donor: Thianthrene Bis(Nitronyl Nitroxide). *J. Am. Chem. Soc.* **2000**, *122* (13), 3234–3235. https://doi.org/10.1021/ja9916759.

(17) Sakurai, H.; Izuoka, A.; Sugawara, T. Design, Preparation, and Electronic Structure of High-Spin Cation Diradicals Derived from Amine-Based Spin-Polarized Donors. *J. Am. Chem. Soc.* **2000**, *122* (40), 9723–9734. https://doi.org/10.1021/ja994547t.

(18) Nakazaki, J.; Chung, I.; Matsushita, M. M.; Sugawara, T.; Watanabe, R.; Izuoka, A.; Kawada, Y. Design and Preparation of Pyrrole-Based Spin-Polarized Donors. *J. Mater. Chem.* **2003**, *13* (5), 1011–1022. https://doi.org/10.1039/B211986B.

(19) Komatsu, H.; Mogi, R.; Matsushita, M. M.; Miyagi, T.; Kawada, Y.; Sugawara, T. Synthesis and Properties of TSF-Based Spin-Polarized Donor. *Polyhedron* **2009**, *28* (9), 1996–2000. https://doi.org/10.1016/j.poly.2008.12.005.

(20) Kusamoto, T.; Kume, S.; Nishihara, H. Realization of SOMO−HOMO Level Conversion for a TEMPO-Dithiolate Ligand by Coordination to Platinum(II). *J. Am. Chem. Soc.* **2008**, *130* (42), 13844–13845. https://doi.org/10.1021/ja805751h.

(21) Kusamoto, T.; Kume, S.; Nishihara, H. Cyclization of TEMPO Radicals Bound to Metalladithiolene Induced by SOMO–HOMO Energy-Level Conversion. *Angew. Chem. Int. Ed.* **2010**, *49* (3), 529–531. https://doi.org/10.1002/anie.200905132.

(22) Kobayashi, Y.; Yoshioka, M.; Saigo, K.; Hashizume, D.; Ogura, T. Hydrogen-Bonding-Assisted Self-Doping in Tetrathiafulvalene (TTF) Conductor. *J. Am. Chem. Soc.* **2009**, *131* (29), 9995–10002. https://doi.org/10.1021/ja809425b.
(23) Westcott, B. L.; Gruhn, N. E.; Michelsen, L. J.; Lichtenberger, D. L. Experimental Observation of Non-Aufbau Behavior: Photoelectron Spectra of Vanadyloctaethylporphyrinate and Vanadylphthalocyanine. *J. Am. Chem. Soc.* **2000**, *122* (33), 8083–8084. https://doi.org/10.1021/ja994018p.
(24) Gryn'ova, G.; Marshall, D. L.; Blanksby, S. J.; Coote, M. L. Switching Radical Stability by pH-Induced Orbital Conversion. *Nat. Chem.* **2013**, *5* (6), 474–481. https://doi.org/10.1038/nchem.1625.
(25) So, S.; Kirk, B. B.; Wille, U.; Trevitt, A. J.; Blanksby, S. J.; Silva, G. da. Reactions of a Distonic Peroxyl Radical Anion Influenced by SOMO–HOMO Conversion: An Example of Anion-Directed Channel Switching. *Phys. Chem. Chem. Phys.* **2020**, *22* (4), 2130–2141. https://doi.org/10.1039/C9CP05989J.
(26) Wang, Y.; Zhang, H.; Pink, M.; Olankitwanit, A.; Rajca, S.; Rajca, A. Radical Cation and Neutral Radical of Aza-Thia[7]Helicene with SOMO–HOMO Energy Level Inversion. *J. Am. Chem. Soc.* **2016**, *138* (23), 7298–7304. https://doi.org/10.1021/jacs.6b01498.
(27) Rajca, A.; Shu, C.; Zhang, H.; Zhang, S.; Wang, H.; Rajca, S. Thiophene-Based Double Helices: Radical Cations with SOMO–HOMO Energy Level Inversion. *Photochem. Photobiol.* **2021**, *97* (6), 1376–1390. https://doi.org/10.1111/php.13475.
(28) Kasemthaveechok, S.; Abella, L.; Jean, M.; Cordier, M.; Roisnel, T.; Vanthuyne, N.; Guizouarn, T.; Cador, O.; Autschbach, J.; Crassous, J.; Favereau, L. Axially and Helically Chiral Cationic Radical Bicarbazoles: SOMO–HOMO Level Inversion and Chirality Impact on the Stability of Mono- and Diradical Cations. *J. Am. Chem. Soc.* **2020**, *142* (48), 20409–20418. https://doi.org/10.1021/jacs.0c08948.
(29) Tanushi, A.; Kimura, S.; Kusamoto, T.; Tominaga, M.; Kitagawa, Y.; Nakano, M.; Nishihara, H. NIR Emission and Acid-Induced Intramolecular Electron Transfer Derived from a SOMO–HOMO Converted Non-Aufbau Electronic Structure. *J. Phys. Chem. C* **2019**, *123* (7), 4417–4423. https://doi.org/10.1021/acs.jpcc.8b08574.
(30) Medina Rivero, S.; Shang, R.; Hamada, Y.; Yan, Q.; Tsuji, H.; Nakamura, E.; Casado, J. Non-Aufbau Spiro-Conjugated Quinoidal & Aromatic Charged Radicals. *Bull. Chem. Soc. Jpn.* **2021**, *94* (3), 989–996. https://doi.org/10.1246/bcsj.20200385.
(31) Sentyurin, V. V.; Levitskiy, O. A.; Bogdanov, A. V.; Yankova, T. S.; Dorofeev, S. G.; Lyssenko, K. A.; Gontcharenko, V. E.; Magdesieva, T. V. Stable Spiro-Fused Diarylaminyl Radicals: A New Type of a Neutral Mixed-Valence System. *Chem. Eur. J.* **2023**, *29* (43), e202301250. https://doi.org/10.1002/chem.202301250.
(32) Kumar, A.; Sevilla, M. D. Proton Transfer Induced SOMO-to-HOMO Level Switching in One-Electron Oxidized A-T and G-C Base Pairs: A Density Functional Theory Study. *J. Phys. Chem. B* **2014**, *118* (20), 5453–5458. https://doi.org/10.1021/jp5028004.
(33) Kumar, A.; Sevilla, M. D. SOMO–HOMO Level Inversion in Biologically Important Radicals. *J. Phys. Chem. B* **2018**, *122* (1), 98–105. https://doi.org/10.1021/acs.jpcb.7b10002.
(34) Murata, R.; Wang, Z.; Miyazawa, Y.; Antol, I.; Yamago, S.; Abe, M. SOMO–HOMO Conversion in Triplet Carbenes. *Org. Lett.* **2021**, *23* (13), 4955–4959. https://doi.org/10.1021/acs.orglett.1c01137.
(35) Wang, Z.; Murata, R.; Abe, M. SOMO–HOMO Conversion in Triplet Cyclopentane-1,3-Diyl Diradicals. *ACS Omega* **2021**, *6* (35), 22773–22779. https://doi.org/10.1021/acsomega.1c03125.
(36) Tsuchiya, T.; Katsuoka, Y.; Yoza, K.; Sato, H.; Mazaki, Y. Stereochemistry, Stereodynamics, and Redox and Complexation Behaviors of 2,2′-Diaryl-1,1′-Biazulenes. *ChemPlusChem* **2019**, *84* (11), 1659–1667. https://doi.org/10.1002/cplu.201900262.
(37) Sabuj, M. A.; Muoh, O.; Huda, M. M.; Rai, N. Non-Aufbau Orbital Ordering and Spin Density Modulation in High-Spin Donor–Acceptor Conjugated Polymers. *Phys. Chem. Chem. Phys.* **2022**, *24* (38), 23699–23711. https://doi.org/10.1039/D2CP02355E.
(38) Abella, L.; Crassous, J.; Favereau, L.; Autschbach, J. Why Is the Energy of the Singly Occupied Orbital in Some Radicals below the Highest Occupied Orbital Energy? *Chem. Mater.* **2021**, *33* (10), 3678–3691. https://doi.org/10.1021/acs.chemmater.1c00683.
(39) Shu, C.; Zhang, H.; Olankitwanit, A.; Rajca, S.; Rajca, A. High-Spin Diradical Dication of Chiral π-Conjugated Double Helical Molecule. *J. Am. Chem. Soc.* **2019**, *141* (43), 17287–17294. https://doi.org/10.1021/jacs.9b08711.
(40) Albrecht, F.; Fatayer, S.; Pozo, I.; Tavernelli, I.; Repp, J.; Peña, D.; Gross, L. Selectivity in Single-Molecule Reactions by Tip-Induced Redox Chemistry. *Science* **2022**, *377* (6603), 298–301. https://doi.org/10.1126/science.abo6471.
(41) Sun, Q.; Mateo, L. M.; Robles, R.; Lorente, N.; Ruffieux, P.; Bottari, G.; Torres, T.; Fasel, R. Bottom-up Fabrication and Atomic-Scale Characterization of Triply Linked, Laterally π-Extended Porphyrin Nanotapes. *Angew. Chem. Int. Ed.* **2021**, *60* (29), 16208–16214. https://doi.org/10.1002/anie.202105350.
(42) Tschitschibabin, A. E. Über Einige Phenylierte Derivate Des *p, p*-Ditolyls. *Ber. Dtsch. Chem. Ges.* **1907**, *40* (2), 1810–1819. https://doi.org/10.1002/cber.19070400282.
(43) Montgomery, L. K.; Huffman, J. C.; Jurczak, E. A.; Grendze, M. P. The Molecular Structures of Thiele's and Chichibabin's Hydrocarbons. *J. Am. Chem. Soc.* **1986**, *108* (19), 6004–6011. https://doi.org/10.1021/ja00279a056.
(44) Yuan, B.; Li, C.; Zhao, Y.; Gröning, O.; Zhou, X.; Zhang, P.; Guan, D.; Li, Y.; Zheng, H.; Liu, C.; Mai, Y.; Liu, P.; Ji, W.; Jia, J.; Wang, S. Resolving Quinoid Structure in Poly(Para-Phenylene) Chains. *J. Am. Chem. Soc.* **2020**, *142* (22), 10034–10041. https://doi.org/10.1021/jacs.0c01930.
(45) Fernández-Rossier, J.; Palacios, J. J. Magnetism in Graphene Nanoislands. *Phys. Rev. Lett.* **2007**, *99* (17), 177204. https://doi.org/10.1103/PhysRevLett.99.177204.

(46) Ortiz, R.; Boto, R. A.; García-Martínez, N.; Sancho-García, J. C.; Melle-Franco, M.; Fernández-Rossier, J. Exchange Rules for Diradical π-Conjugated Hydrocarbons. *Nano Lett.* **2019**, *19* (9), 5991–5997. https://doi.org/10.1021/acs.nanolett.9b01773.

(47) Mishra, S.; Vilas-Varela, M.; Lieske, L.-A.; Ortiz, R.; Fatayer, S.; Rončević, I.; Albrecht, F.; Frederiksen, T.; Peña, D.; Gross, L. Bistability between *π*-Diradical Open-Shell and Closed-Shell States in Indeno[1,2-*a*]Fluorene. *Nat. Chem.* **2024**, 1–7. https://doi.org/10.1038/s41557-023-01431-7.

(48) Bedi, A.; Shimon, L. J. W.; Gidron, O. Helically Locked Tethered Twistacenes. *J. Am. Chem. Soc.* **2018**, *140* (26), 8086–8090. https://doi.org/10.1021/jacs.8b04447.

(49) Uhlmann, C.; Swart, I.; Repp, J. Controlling the Orbital Sequence in Individual Cu-Phthalocyanine Molecules. *Nano Lett.* **2013**, *13* (2), 777–780. https://doi.org/10.1021/nl304483h.

(50) Kim, W. Y.; Kim, K. S. Tuning Molecular Orbitals in Molecular Electronics and Spintronics. *Acc. Chem. Res.* **2010**, *43* (1), 111–120. https://doi.org/10.1021/ar900156u.

# Supporting Information

# Observation of SOMO-HOMO inversion in a polycyclic conjugated hydrocarbon

Shantanu Mishra[1], Manuel Vilas-Varela[2], Shadi Fatayer[3], Florian Albrecht[1], Diego Peña[2,4], and Leo Gross[1]

[1]IBM Research Europe – Zurich, 8803 Rüschlikon, Switzerland.

[2]Center for Research in Biological Chemistry and Molecular Materials (CiQUS) and Department of Organic Chemistry, University of Santiago de Compostela, 15782 Santiago de Compostela, Spain.

[3]Applied Physics Program, Physical Science and Engineering Division, King Abdullah University of Science and Technology (KAUST), 23955-6900 Thuwal, Kingdom of Saudi Arabia.

[4]Oportunius, Galician Innovation Agency (GAIN), 15702 Santiago de Compostela, Spain.

Corresponding authors: Shantanu Mishra (SHM@zurich.ibm.com), Diego Peña (diego.pena@usc.es) and Leo Gross (LGR@zurich.ibm.com).

# 1. Scanning probe microscopy measurements and calculations.

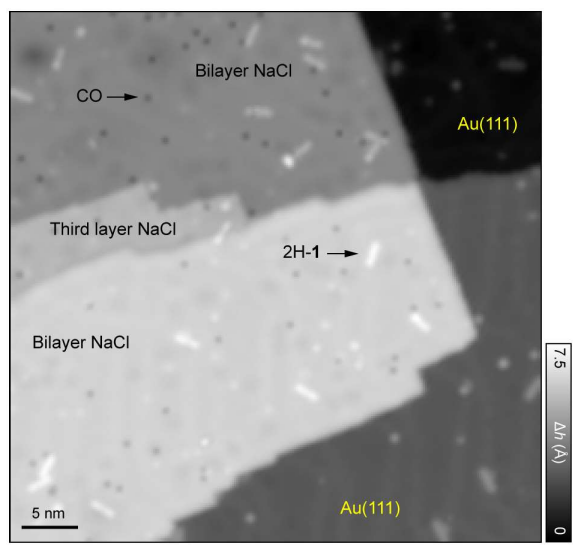

**Figure S1.** Overview STM image of the sample acquired with a gold-terminated tip ($V$ = 0.2 V and $I$ = 0.3 pA). The sample consists of large bilayer NaCl islands on Au(111), a minority of third-layer NaCl islands, carbon monoxide (CO) molecules that appear as circular depressions, and 2H-**1** molecules.

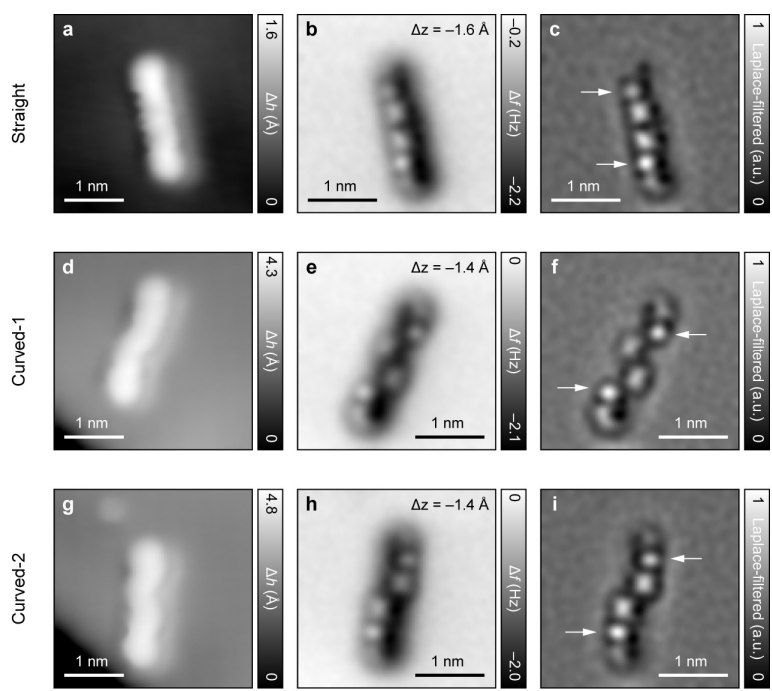

**Figure S2.** Structural characterization of 2H-**1**. On bilayer NaCl/Au(111), 2H-**1** is found in three different geometries, namely, straight, curved-1 and curved-2, which differ in the relative tilting of the hexagonal rings of the central biphenyl moiety and the terminal indene units. (a, d, g) In-gap STM images of straight (a), curved-1 (d) and curved-2 (g) species ($V$ = 0.2 V and $I$ = 0.5 pA). The molecules in (d) and (g) are adsorbed close to a NaCl-Au step edge. (b, e, h) Corresponding AFM images of straight (b), curved-1 (e) and curved-2 (h) species. STM setpoint for AFM images: $V$ = 0.2 V and $I$ = 0.5 pA on bilayer NaCl. (c, f, i) Laplace-filtered AFM images of straight (c), curved-1 (f) and curved-2 (i) species. The arrows in the Laplace-filtered AFM images indicate the bright protrusions corresponding to -CH$_2$- moieties in the indene units.

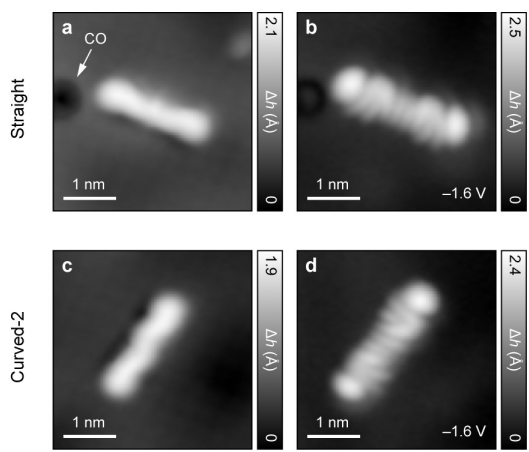

**Figure S3.** Orbital density imaging of 2H-**1**. (a, c) In-gap STM images of straight (a) and curved-2 (c) species ($V$ = 0.2 V, $I$ = 0.5 pA). A co-adsorbed carbon monoxide (CO) molecule next to 2H-**1** is indicated in (a). (b, d) Corresponding STM images acquired at the positive ion resonance ($I$ = 0.2 pA), revealing the HOMO density of both molecules. It is noted that the onset of the positive ion resonances of **1**, 1H-**1** and 2H-**1** are all at $V$ = –1.5 V. We could not detect the negative ion resonance of 2H-**1** up to $V$ = 2.8 V. Above this bias voltage, 2H-**1** exhibited high mobility and it was not possible to image the molecule stably.

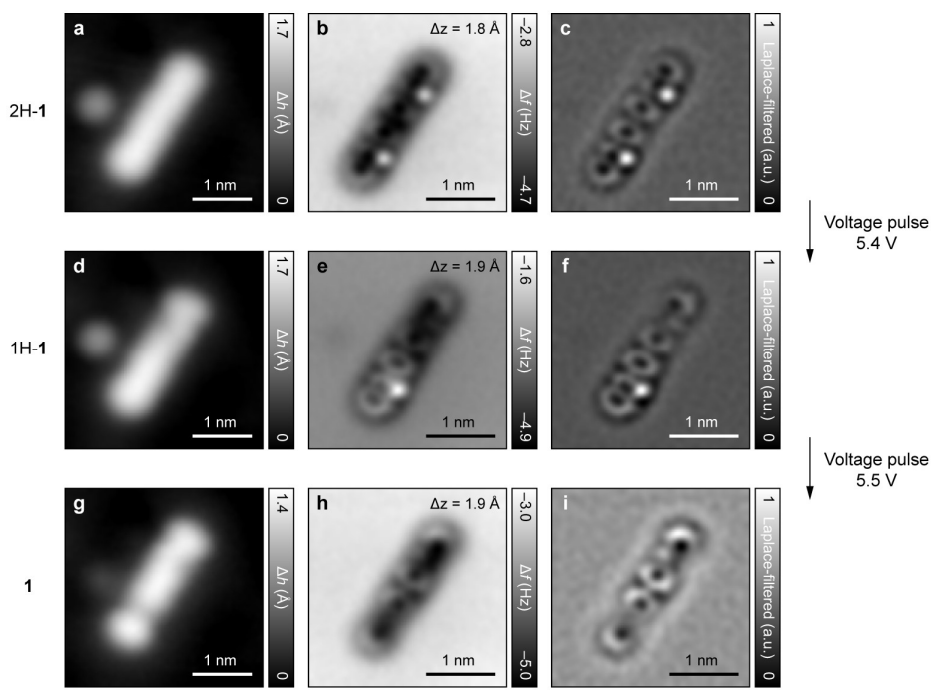

**Figure S4.** Generation of **1** on Au(111). To generate **1** on Au(111), the STM tip was located at the center of 2H-**1** at $V$ = 0.2 V and $I$ = 0.3 pA, and the feedback loop was switched off. The tip was then retracted by 5.0–5.5 Å to limit the tunneling current to ~10 pA (at the maximum applied bias voltage), and the bias voltage was ramped to 5.4–5.6 V. Abrupt changes in the tunneling current indicated possible dehydrogenation events, and the molecule was subsequently scanned to monitor any changes. (a, d, g) STM images showing stepwise generation of **1** from 2H-**1** ($V$ = 0.2 V and $I$ = 0.3 pA). Starting from 2H-**1** (a), an initial voltage pulse of 5.4 V led to the generation of 1H-**1** (d), and a further voltage pulse of 5.5 V resulted in the generation of **1** (g). The round protrusion next to the molecule in the STM images is an unidentified adsorbate. (b, e, h) Corresponding AFM images of 2H-**1** (b), 1H-**1** (e) and **1** (h). STM setpoint for AFM images: $V$ = 0.2 V and $I$ = 0.3 pA on Au(111). (c, f, i) Laplace-filtered AFM images of 2H-**1** (c), 1H-**1** (f) and **1** (i). The hexagonal rings of the biphenyl moiety of 2H-**1** are tilted to a much lesser degree for the molecule on Au(111) than on bilayer NaCl (compare with Figure S1). Furthermore, we could not detect any electronic-state signatures of **1** on Au(111) likely due to a strong molecule-metal interaction.

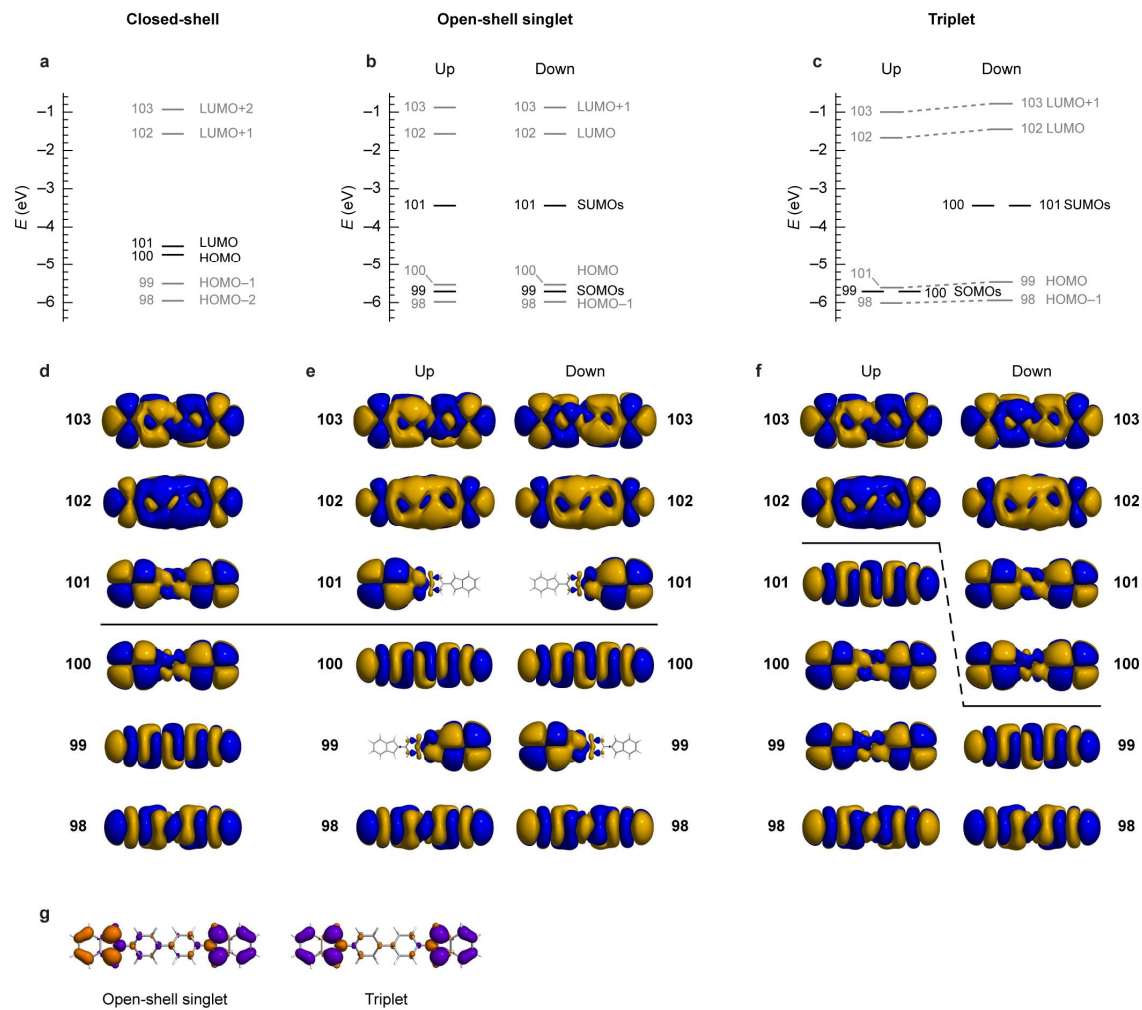

**Figure S5.** Gas-phase density functional theory (DFT) calculations of **1**. (a–c) DFT energy spectrum of **1** in the closed-shell (a), open-shell singlet (b) and triplet (c) states. The open-shell singlet and triplet states are energetically degenerate – the singlet-triplet gap is < 1 meV, which indicates negligible electronic coupling between the indenyl units. The closed-shell state is ~1 eV higher in energy than the open-shell singlet and triplet states. The orbitals for each spin level are numbered in the order of increasing energy. (d–f) Corresponding DFT-calculated wavefunctions of the orbitals of **1** in the closed-shell (d), open-shell singlet (e) and triplet (f) states (isovalue: 0.001 e$^-$/Å$^3$). Orbitals above and below the solid lines are unoccupied and occupied, respectively. (g) DFT-calculated spin density of **1** in the open-shell singlet and triplet states (isovalue: 0.05 e$^-$/Å$^3$). The DFT results qualitatively agree with the tight-binding and mean-field Hubbard calculations, and SHI is observed for **1** in the open-shell singlet and triplet states.

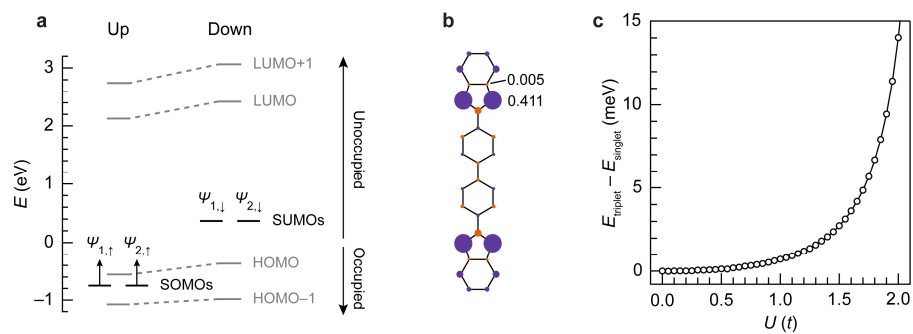

**Figure S6.** Mean-field Hubbard calculations of **1**. (a) Mean-field Hubbard energy spectrum of **1** in the triplet state. Zero of the energy axis is set to lie in between the HOMO and SUMOs. States above and below zero energy are unoccupied and occupied, respectively. The nearest-neighbor hopping parameter $t$ = 2.7 eV, and the on-site Coulomb repulsion $U$ ~4 eV (that is, $1.5 \times t$). As for the open-shell singlet state of **1**, SHI is present for the triplet state as well. (b) Spin-polarization plot of **1** in the triplet state. The numbers denote absolute values of spin polarization. (c) The singlet-triplet gap, expressed as the difference in energies between the triplet and singlet states ($E_{triplet} - E_{singlet}$), as a function of $U$. Here, $U$ is expressed in units of $t$. While the open-shell singlet state is lower in energy than the triplet state for all values of $U$, the calculated singlet-triplet gap is < 5 meV for $U \leq 1.7 \times t$.

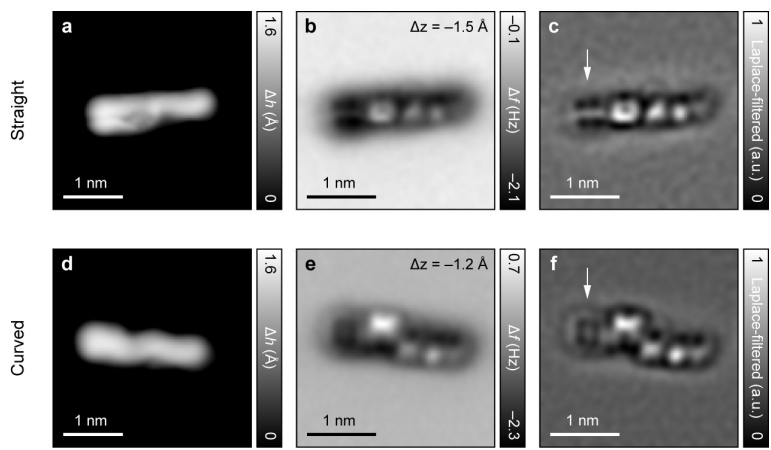

**Figure S7.** Structural characterization of 1H-**1**. On bilayer NaCl/Au(111), 1H-**1** is found in two different geometries, namely, straight and curved, which differ in the relative tilting of the hexagonal rings of the central biphenyl moiety. The rings tilt in the same (opposite) direction in straight (curved) species. (a, d) In-gap STM images of straight (a) and curved (d) species ($V$ = 0.2 V and $I$ = 0.5 pA). Panels (a) and (d) show the same molecule, wherein movement of the molecule resulted in the switching of its geometry. (b, e) Corresponding AFM images of straight (b) and curved (e) species. STM setpoint for AFM images: $V$ = 0.2 V and $I$ = 0.5 pA on bilayer NaCl. (c, f) Laplace-filtered AFM images of straight (c) and curved (f) species. Similar to **1**, 1H-**1** on bilayer NaCl/Au(111) exhibits movement during scanning. The arrows in the Laplace-filtered AFM images indicate the indenyl unit, which appears bisected due to movement of the molecule between different adsorption sites under the influence of the tip.

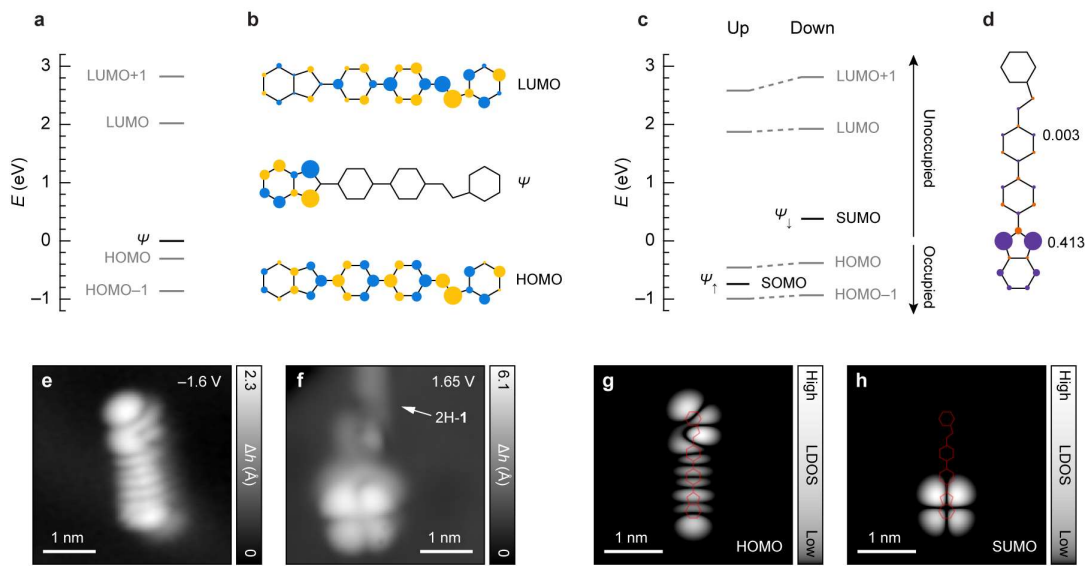

**Figure S8.** Electronic characterization of 1H-**1**. In tight-binding calculations, the structure of 1H-**1** is approximated by removing the carbon atom belonging to the -CH$_2$- group from the molecular framework. This approach is justified by this $sp^3$-hybridized carbon atom being out of conjugation with the rest of the π-system. (a) Tight-binding energy spectrum of 1H-**1** ($t$ = 2.7 eV). Zero of the energy axis is set to match the energy of the state $\Psi$ that is localized at the indenyl unit. (b) Tight-binding wavefunctions of HOMO, $\Psi$ and LUMO. (c) Mean-field Hubbard energy spectrum of 1H-**1** in the doublet state ($U$ ~4 eV), revealing SHI. Zero of the energy axis is set to lie in between the HOMO and SUMO. States above and below zero energy are unoccupied and occupied, respectively. (d) Spin-polarization plot of 1H-**1** in the doublet state. The numbers denote absolute values of spin polarization. (e, f) STM images of 1H-**1** acquired at the positive ($I$ = 0.15 pA; e) and negative ($I$ = 0.2 pA; f) ion resonances indicating SHI. The STM images are acquired on two different molecules. The 1H-**1** molecule in (f) is adsorbed next to a 2H-**1** species that is indicated by the arrow. (g, h) Constant-height mean-field Hubbard local density of states maps of HOMO (g) and SUMO (h) of 1H-**1**, calculated at a height of 5.6 Å above the molecular plane and shown in a logarithmic color scale.

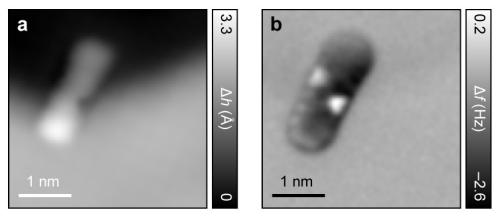

**Figure S9.** (a) In-gap STM image of **1** whose orbital density images are shown in Figure 3 ($V$ = 0.2 V and $I$ = 0.5 pA). (b) Constant-current AFM image acquired simultaneously with (a). The two bright features at the center of **1** in (b) correspond to the strongly-tilted hexagonal rings of the biphenyl moiety.

## 2. Solution synthesis and mass spectrometry of 2H-**1**.

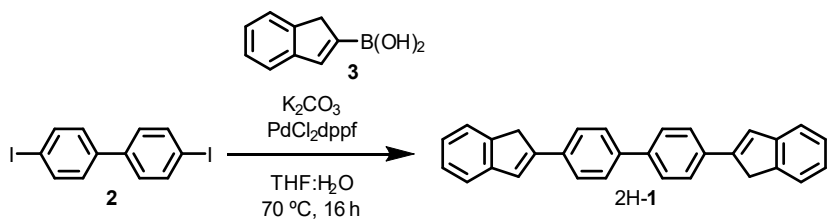

**Scheme S1.** Solution synthesis of 2H-**1**.

**Solution synthesis of 4,4′-di(1H-inden-2-yl)-1,1′-biphenyl (2H-1).** Over a degassed mixture of **2** (100 mg, 0.25 mmol), boronic acid **3** (100 mg, 0.66 mmol) and $K_2CO_3$ (170 mg, 1.25 mmol) in THF:$H_2O$ (4:1, 10 mL), $PdCl_2dppf$ (10 mg, 0.01 mmol) was added. The resulting mixture was heated at 70 °C for 16 h. Then, the reaction mixture was cooled to room temperature and filtered. The precipitate was then washed with $H_2O$ (2 × 10 mL), MeOH (2 × 10 mL), acetone (2 × 10 mL), $Et_2O$ (2 × 10 mL) and $CH_2Cl_2$ (2 × 10 mL), leading to 2H-**1** (30 mg, 31% yield) as an extremely insoluble solid. High-resolution mass spectra (atmospheric pressure chemical ionization method): calculated 382.1716, experimental 382.1728.

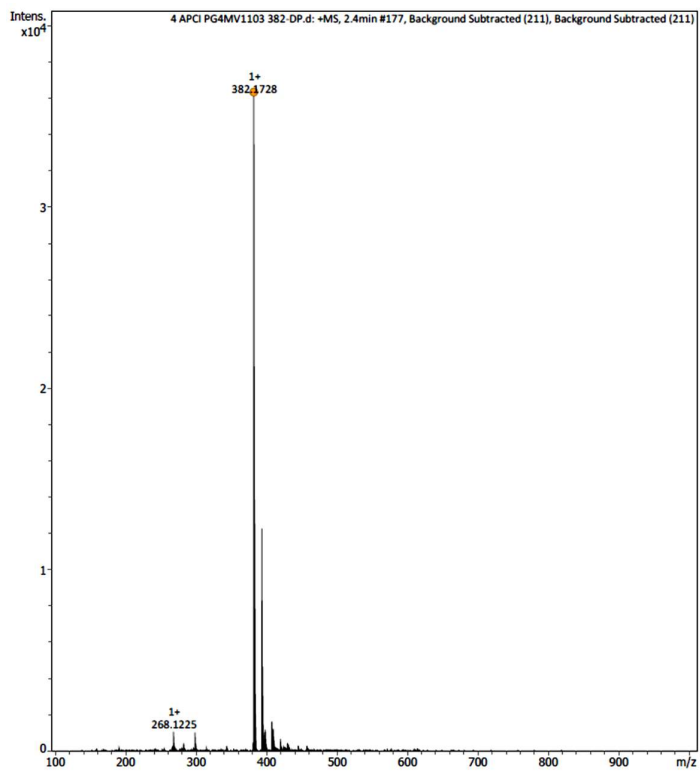

**Figure S10.** Mass spectrum of 2H-**1**.

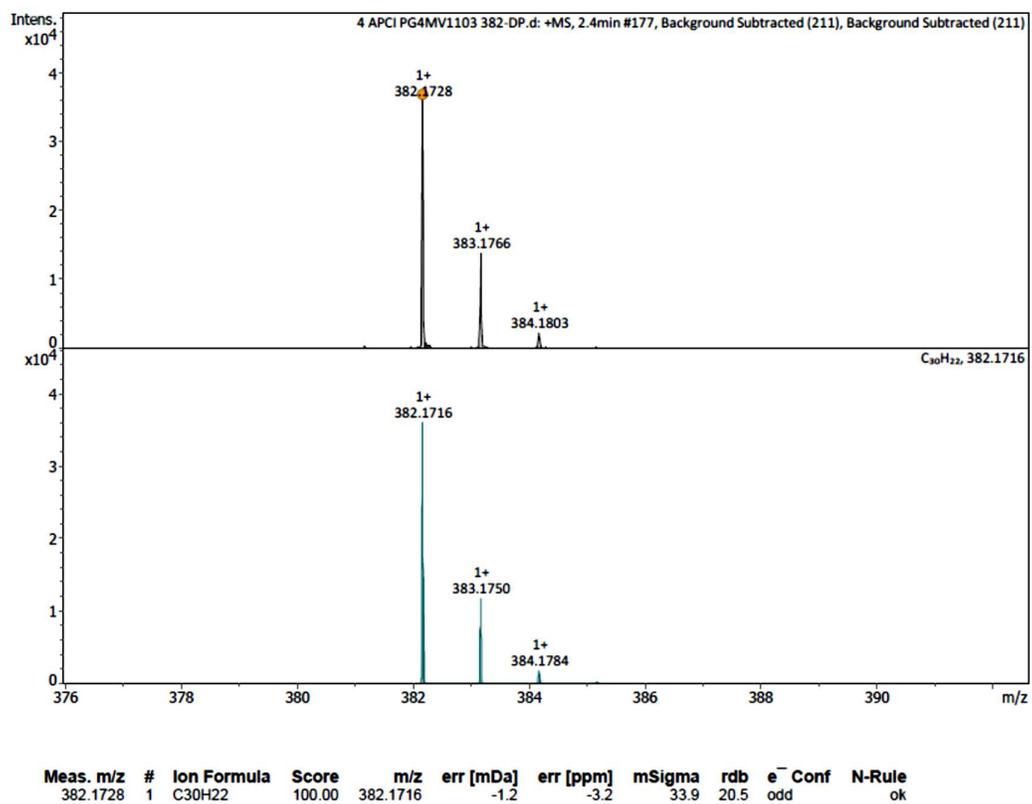

**Figure S11.** Experimental (top) and simulated (bottom) high-resolution mass spectra of 2H-**1**.

## 3. Experimental and theoretical methods.

**1. General chemical procedures.** Starting materials were purchased reagent grade from ABCR and Sigma-Aldrich, and used without further purification. Reactions were carried out in flame-dried glassware and under an inert atmosphere of purified argon using Schlenk techniques. Thin-layer chromatography was performed on silica gel 60 F-254 plates (Merck). Column chromatography was performed on silica gel (40-60 μm). Mass spectrometry and high-resolution mass spectrometry data were measured on a Bruker microTOF spectrometer.

**2. Sample preparation and scanning probe microscopy measurements.** Au(111) surface was prepared by multiple cycles of sputtering with Ne$^+$ ions and annealing up to 800 K. NaCl was thermally evaporated on Au(111) surface held at 323 K, which resulted in the growth of predominantly bilayer (100)-terminated islands, with a minority of third-layer islands. Submonolayer coverage of 2H-**1** on the surface was obtained by flashing an oxidized silicon wafer containing 2H-**1** molecules in front of the cold sample in the microscope. Carbon monoxide molecules for tip functionalization were dosed from the gas phase on the cold sample. STM and AFM measurements were performed in a home-built system operating at base pressures below $1 \times 10^{-10}$ mbar and a base temperature of 5 K. Bias voltages are provided with respect to the sample. Unless otherwise mentioned, all STM and AFM measurements were performed with carbon-monoxide-functionalized tips. AFM measurements were performed in non-contact mode with a qPlus sensor.[1] The sensor was operated in frequency-modulation mode[2] with a constant oscillation amplitude of 0.5 Å. Unless noted otherwise, STM measurements were performed in constant-current mode, and AFM measurements were performed in constant-height mode with $V = 0$ V. STM and AFM images were post-processed using Gaussian low-pass filters.

**3. Tight-binding calculations.** Tight-binding and mean-field Hubbard calculations were performed by numerically solving the following mean-field Hubbard Hamiltonian with nearest-neighbor hopping

$$\widehat{H}_{MFH} = -t \sum_{\langle i,j \rangle, \sigma} c_{i,\sigma}^{\dagger} c_{j,\sigma} + U \sum_{i,\sigma} \langle n_{i,\sigma} \rangle n_{i,\bar{\sigma}} - U \sum_{i} \langle n_{i,\uparrow} \rangle \langle n_{i,\downarrow} \rangle. \quad (1)$$

Here, $c_{i,\sigma}^{\dagger}$ and $c_{j,\sigma}$ denote the spin selective ($\sigma \in \{\uparrow, \downarrow\}$ with $\bar{\sigma} \in \{\downarrow, \uparrow\}$) creation and annihilation operator at neighboring sites $i$ and $j$, $t$ = 2.7 eV is the nearest-neighbor hopping parameter, $U$ is the on-site Coulomb repulsion, $n_{i,\sigma}$ and $\langle n_{i,\sigma} \rangle$ denote the number operator and mean occupation number at site $i$, respectively. Orbital electron densities, $\rho$, of the $n^{\text{th}}$-eigenstate with energy $E_n$ have been simulated from the corresponding state vector $a_{n,i,\sigma}$ by

$$\rho_{n,\sigma}(\vec{r}) = \left| \sum_{i} a_{n,i,\sigma} \phi_{2p_z}(\vec{r} - \vec{r}_i) \right|^2, \quad (2)$$

where $\phi_{2p_z}$ denotes the Slater 2$p_z$ orbital for carbon.

**4. Density functional theory calculations.** DFT calculations were performed using the FHI-aims code.[3] Closed-shell, open-shell singlet and triplet states of **1** were independently investigated in the gas phase. The B3LYP exchange-correlation functional, using the Vosko-Wilk-Nusair local-density approximation as implemented in the FHI-aims code, was employed.[4] We used the van der Waals scheme by Tkatchenko and Scheffler.[5] The convergence criteria for all calculations were set to 10$^{-3}$ eV/Å for the total forces and 10$^{-5}$ eV for the total energies. Open-shell calculations were performed in the spin-polarized (unrestricted) framework and closed-shell calculations were performed in the non-spin-polarized (restricted) framework. The basis defaults were set to "really tight" for all elements. For open-shell calculations, an initial spin moment was placed at the two apical carbon atoms of each of the pentagonal rings.

## 4. References.

(1) Giessibl, F. J. High-Speed Force Sensor for Force Microscopy and Profilometry Utilizing a Quartz Tuning Fork. *Appl. Phys. Lett.* **1998**, *73* (26), 3956–3958. https://doi.org/10.1063/1.122948.
(2) Albrecht, T. R.; Grütter, P.; Horne, D.; Rugar, D. Frequency Modulation Detection Using High-Q Cantilevers for Enhanced Force Microscope Sensitivity. *J. Appl. Phys.* **1991**, *69* (2), 668–673. https://doi.org/10.1063/1.347347.
(3) Blum, V.; Gehrke, R.; Hanke, F.; Havu, P.; Havu, V.; Ren, X.; Reuter, K.; Scheffler, M. Ab Initio Molecular Simulations with Numeric Atom-Centered Orbitals. *Comput. Phys. Commun.* **2009**, *180* (11), 2175–2196. https://doi.org/10.1016/j.cpc.2009.06.022.
(4) Scuseria, G. E.; Staroverov, V. N. Progress in the Development of Exchange-Correlation Functionals. In *Theory and Applications of Computational Chemistry*; Elsevier, 2005; pp 669–724. https://doi.org/10.1016/B978-044451719-7/50067-6.
(5) Tkatchenko, A.; Scheffler, M. Accurate Molecular Van Der Waals Interactions from Ground-State Electron Density and Free-Atom Reference Data. *Phys. Rev. Lett.* **2009**, *102* (7), 073005. https://doi.org/10.1103/PhysRevLett.102.073005.